\documentclass[prb,twocolumn,showpacs]{revtex4}
\usepackage{bm}
\usepackage{color} 
\usepackage{graphicx}
\usepackage{epsfig}
\usepackage{amsmath}
\usepackage{amssymb}
\usepackage{soul}

\begin{document}

\title{Ratchet transport of a two-dimensional electron gas at cyclotron resonance}

\author{G.\,V.\,Budkin}
\author{S.\,A.\,Tarasenko}

\affiliation{Ioffe Institute, 194021 St.~Petersburg, Russia}


% 73.63.-b Electronic transport in nanoscale materials and
% structures (see also 73.23.-b Electronic transport in mesoscopic
% systems)
% 73.63.Hs Quantum wells
% 73.50.Fq 	High-field and nonlinear effects
% 76.40.+b 	Diamagnetic and cyclotron resonances
% 73.50.Pz Photoconduction and photovoltaic effects  

% 73.50.-h Electronic transport phenomena in thin films
% 73.50.Bk General theory, scattering mechanisms
% 73.50.Jt Galvanomagnetic and other magnetotransport effects
% (including thermomagnetic effects)
% 73.50.Fq High-field and nonlinear effects  
% 73.50.Pz Photoconduction and photovoltaic effects  

\begin{abstract}
 The driving of charge carriers confined in a quantum well lacking the center of space inversion by an alternating electric field leads to the formation of a direct electric current. We develop a microscopic theory of such a quantum ratchet effect for quantum wells subjected to a static magnetic field. We show that the ratchet current emerges for a linearly polarized alternating electric field as well as a rotating electric field and drastically increases at the cyclotron resonance conditions. For the magnetic field tilted with respect to the quantum well normal, the ratchet current contains an additional resonance at the first subharmonic of the cyclotron resonance.    
\end{abstract}

\maketitle

\section{Introduction}

The response of a conducting system to an alternating electric field is one of the central topics of research in solid-state electrodynamics. Besides the linear response, where the induced electric current oscillates at the frequency of the electric field and its amplitude linearly scales with the field amplitude, ac electric field can give rise to a direct motion of charge carriers. Such an electric rectification (or electronic ratchet effect) naturally occurs in macroscopically inhomogeneous structures such as diodes or field-effect transistors,\cite{Dyakonov1996,Knap2009} asymmetric lattices,\cite{Blanter1998,Hoehberger2001,Olbrich2009,Popov2013,Budkin2014,Rozhansky2015} or systems with asymmetric patterning.\cite{Entin2006,Sassine08,Chepelianskii08,Bisotto2011,Koniakhin2014} The ratchet transport of carriers arises also in macroscopically homogenous structures (homogenous in all three dimensions for bulk materials or homogenous in the plane for two-dimensional systems) provided the structures lack the center of space inversion.\cite{Sturman_book,Falko1989,Tarasenko2011,Entin2013,Drexler2013} The ratchet effects are used for the study of spatial symmetry of semiconductor structures,\cite{Belkov2008} details of the electron energy spectrum,\cite{Olbrich2013,Dantscher2015} and also underlie the operation of fast detectors of microwave and terahertz radiation.\cite{Knap2009,Ganichev2007} 

The efficiency of the generation of a dc current by an ac electric field can be considerably enhanced in an external magnetic field 
if the frequency of the ac field is close to the carrier cyclotron frequency. Previously, such an enhancement of the electric response at the cyclotron resonance conditions was demonstrated for the spin currents in HgTe quantum wells (QWs)\cite{Olbrich2013} and HgTe-based three-dimensional topological insulators\cite{Dantscher2015} and for the electric currents of photon drag induced by teraherzt radiation.\cite{Dmitriev1991,Stachel2014} The orbital ratchet transport of electrons in macroscopically homogenous QWs in the presence of a normal magnetic field has not been analyzed so far. 

Here, we develop a quasi-classical microscopic theory of the orbital ratchet effect in QWs subjected to a static magnetic field and show that the arising current drastically increases at the cyclotron resonance. We consider the ratchet current caused by 
the QW structure inversion asymmetry which may originate from an asymmetry in the confinement potential or doping profile, or induced by a gate voltage. Two geometries, at which the ratchet current emerges and which can be experimentally realized, are analyzed: (i) the static magnetic field is normal to the QW plane and the ac electric field has both the in-plane and out-of-plane components and (ii) 
the magnetic field is tilted with respected to the QW normal and the ac electric field is polarized in the QW plane. For the latter geometry we find that the dc electric current has an additional resonant contribution. This resonance occurs when the frequency of the ac field matches the double cyclotron frequency, in the spectral range where the free-carrier absorption has no peculiarity. We also calculate the spatial distribution of the electrostatic potential induced by the ratchet current in infinite and finite-size samples and show that the distribution depends on the magnetic field strength, the sample geometry, and the boundary conditions used. 

\section{Perpendicular magnetic field}\label{Sec_perpB}

\begin{figure}[t]
\includegraphics[scale=0.5]{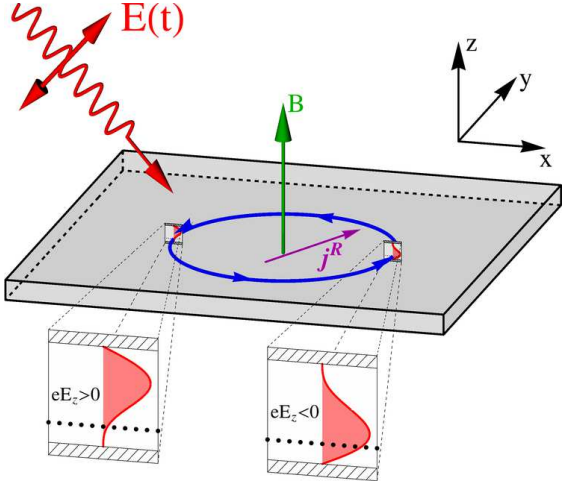}
\caption{\label{geometry_f} 
Microscopic model of the generation of a dc electric current $\bm j^{\rm R}$ by an ac electric field $\bm E(t)$ in an asymmetric QW subjected to a static magnetic field $\bm B$ at the cyclotron resonance conditions. The dc component of the electric current emerges due to the combined action of the in-plane component of the electric field, which induces the cyclotron motion of electrons in the QW plane (blue ellipse), and the field out-of-plane component $E_z$, which causes the electron mobility modulation at the same frequency. Insets show
that the electric force $e E_z$ shifts the electron density to the top (or bottom) interface thereby decreasing (or increasing) the electron scattering by impurities (black dots).}
\end{figure}

We begin the study with the geometry of the static magnetic field $\bm B$ pointing along the 
QW normal $z$, see Fig.~\ref{geometry_f}. In this configuration, a dc electric current $\bm j^{\rm R}$ emerges if the ac electric field
\begin{equation}
\bm{E}(t) = \bm{E} e^{-i \omega t}+\bm{E}^* e^{i \omega t} \:,
\end{equation}
where $\bm{E}$ and $\omega$ are the field amplitude and frequency, has both the in-plane $\bm E_{\parallel}$ and out-of-plane $E_z$ components.

Figure~\ref{geometry_f} illustrates the microscopic mechanism of the dc current formation. The in-plane component  $\bm E_{\parallel}(t)$ of the electric field together with the static magnetic field $\bm B$ causes the motion of electrons in elliptical orbits in the QW plane
at the field frequency $\omega$. Synchronously with this in-plane motions, the electric field out-of-plane component $E_z(t)$ pushes the electrons to the top or bottom interfaces of the QW depending on the field polarity. The corresponding distributions of the electron density in the QW cross section for positive and negative $e E_z$, where $e$ is the electron charge, are sketched in the insets. The shift of the electron density along the $z$ axis in the asymmetric QW results, in turn, in the modulation of the electron mobility at the field frequency $\omega$. In the insets in Fig.~\ref{geometry_f}, the QW asymmetry is modeled by placing the $\delta$-layer of impurities
(black dots), which cause electron scattering and control the electron mobility, closer to the bottom interface. As a result, the oscillating electron motion in the QW plane driven by $\bm E_{\parallel}(t)$ together with the mobility modulation at the same frequency induced by $E_z(t)$ leads to the generation of a direct electric current $\bm j^{\rm R}$. At the cyclotron resonance conditions, the amplitude of the oscillating electron motion in the QW plane increases and so does the efficiency of the dc current generation.

A quasi-classical theory of the orbital ratchet effect described above can be developed in the framework of the Boltzmann transport equation. In this approach, the ac electric field and the static magnetic field are considered as the electric force and the 
Lorentz force, respectively, acting upon the electrons. The approach is relevant if the energy $\hbar \omega$ is much smaller 
than the mean kinetic energy of the electrons and the magnetic field is within the classical range. The electron distribution in the momentum space is described by the time-dependent function $f_{\bm p}(t)$ which satisfies the Boltzmann equation
\begin{equation}
 \label{kinetic_equation}
 \dfrac{\partial f_{\bm p} }{\partial t} + e \left(\bm{E}_{\|} (t) + \frac{1}{c} [ \bm{v} \times \bm{B}] \right) \dfrac{\partial f_{\bm p}}{\partial \bm{p}} = {\rm St}  f_{\bm p} \:, 
\end{equation}
where $\bm p$ is the momentum, $\bm v = d \varepsilon_{p} / d \bm p$ and $\varepsilon_{p}$ are the electron velocity and energy,
respectively, and ${\rm St}  f_{\bm p}$ is the collision integral. For elastic scattering, the collision integral has the form
\begin{equation}\label{St}
{\rm St}  f_{\bm p} = \sum \limits_{\bm{p}'} \left( W_{\bm{p} \bm{p}'} f_{\bm{p}'} - W_{\bm{p}' \bm{p}} f_{\bm{p}} \right) \:,
\end{equation}
where $W_{\bm{p} \bm{p}'}$ is the rate of scattering.  

The rate of electron scattering in an asymmetric QW to first order in $E_z$ can be presented in the form 
\begin{equation}\label{W_pp}
W_{\bm{p} \bm{p}'} = W_{\bm{p} \bm{p}'}^{(0)} + e E_z(t) \: W_{\bm{p} \bm{p}'}^{(1)} \:,
\end{equation}
where $W_{\bm{p} \bm{p}'}^{(0)}$ is the scattering rate at zero electric field and $W_{\bm{p} \bm{p}'}^{(1)}$ is the field-induced correction. In the first Born approximation, the term $W_{\bm{p} \bm{p}'}^{(0)}$ is given by
\begin{equation}
W_{\bm{p} \bm{p}'}^{(0)} = \frac{2 \pi}{\hbar} \langle |V_{11}(\bm p,\bm p')|^2 \rangle \, \delta (\varepsilon_p -\varepsilon_{p'}) \:,
\end{equation}
where $V_{11}(\bm p,\bm p')$ is the matrix element of intrasubband scattering between the states with the momenta $\bm p$ and $\bm p'$ and 
the angle brackets denote averaging over the positions of scattering centers. The second term in the right-hand side of Eq.~\eqref{W_pp} originates from the admixture of excited-subband states to the ground-subband states by the out-of-plane electric field and is given by~\cite{Tarasenko2011} 
\begin{equation}
W_{\bm{p} \bm{p}'}^{(1)} = \dfrac{8 \pi}{\hbar}\sum\limits_{\nu \neq 1 } \dfrac{z_{\nu 1}}{\varepsilon_{ \nu 1}} \langle \operatorname{Re} V^*_{11}(\bm p, \bm p') V_{1\nu} (\bm p, \bm p') \rangle \, \delta (\varepsilon_p -\varepsilon_{p'})\:,
\end{equation}
where $\nu$ is the index of the excited electron subbands, $z_{\nu 1}$ is the intersubband matrix element of the coordinate operator,
$\varepsilon_{\nu 1}$ is the energy separation between the subbands, and $V_{1\nu} (\bm p, \bm p')$ is the matrix element of intersubband scattering. We note that the use of the collision integral~\eqref{St} with the time-dependent scattering rate~\eqref{W_pp} is justified
in the adiabatic regime, when $\hbar \omega$ is much smaller than the energy separation between the excited and ground electron subbands.

To solve the Boltzmann equation~\eqref{kinetic_equation} we decompose the distribution function $f_{\bm p}(t)$ in the Fourier series of frequency and angular harmonics as follows
\begin{equation}
\label{distibution_decomposition}
f_{\bm p}(t) = \sum\limits_{n,m} f^{n,m}(p) \exp(i m \varphi_{\bm p} - i n \omega t) \:,
\end{equation}
where $\varphi_{\bm p} = \arctan (p_y/p_x)$ is the polar angle of $\bm p$. Accordingly, the collision integral for the scattering rate $W_{\bm{p} \bm{p}'}$ dependent of $|\varphi_{\bm p} - \varphi_{\bm p'}|$ is rewritten in the form
\begin{align}\label{collision_integral}
{\rm St}  f_{\bm p} = & - \sum_n \sum_{m \neq 0} \left[ \dfrac{f^{n,m}}{\tau_m}  + e \zeta_m \left( E_z f^{n-1,m} + E_z^* f^{n+1,m} \right) \right] \nonumber \\
&\times \exp(i m \varphi_{\bm p} - i n \omega t) \:.
\end{align}
Here, $\tau_m$ is the relaxation time of the $m$-th angular harmonic of the electron distribution function, 
\begin{equation}
\tau_m^{-1} = \sum_{\bm{p}'} W_{\bm{p} \bm{p}'}^{(0)}\left[1-\cos m(\varphi_{\bm p} - \varphi_{\bm p'} )\right] \:,
\end{equation}
and 
\begin{equation}
\zeta_m = \sum_{\bm{p}'} W_{\bm{p} \bm{p}'}^{(1)}\left[1-\cos m(\varphi_{\bm p} - \varphi_{\bm p'} )\right] \:.
\end{equation}

In the harmonics representation, the Boltzmann equation~\eqref{kinetic_equation} has the form of the set of linear equations
\begin{multline}\label{linear_set1}
\Gamma^{n,m} f^{n,m}+e \zeta_m \left(E_z f^{n-1,m}+E_z^* f^{n+1,m}\right) (1-\delta_{m,0})\\
+\dfrac{e \bm{E}_{\|}}{2} \cdot  
\left(\bm{o}_{-} \hat{K}^m_{-} f^{n-1,m-1} +
\bm{o}_{+} \hat{K}^m_{+} f^{n-1,m+1}\right)\\
+\dfrac{e \bm{E}_{\|}^*}{2} \cdot 
\left(\bm{o}_{-} \hat{K}^m_{-} f^{n+1,m-1} +
\bm{o}_{+} \hat{K}^m_{+} f^{n+1,m+1}\right)=0\:,
\end{multline}
where $\Gamma^{n,m} = 1/\tau_m - i n \omega - i m \omega_c$, $\omega_c=e B_z / (m_c c)$ is the cyclotron frequency, $m_c = p/v$ is the cyclotron mass, $\bm{o}_{\pm}=\bm{o}_x \pm i \bm{o}_y$, $\bm{o}_x$ and $\bm{o}_y$ are the unit vectors along the $x$ and $y$ axes, respectively, and $\hat{K}^m_{\pm} = d / d p \pm (m \pm 1)/p$.

In thermal equilibrium, when the ac electric field is absent, the distribution function contains only the harmonic $f^{(0,0)}$
which is given by the Fermi-Dirac distribution $f_0(\varepsilon_p)$. To first order in the electric field amplitude, solution of the equation set~\eqref{linear_set1} has the form
\begin{align}\label{f11}
f^{1,1} = - \frac{e \tau_1 \bm E_{\parallel} \cdot \bm o_-}{2 [1 - i(\omega + \omega_c)\tau_1]} \frac{d f_0(\varepsilon_p)}{d p} \:, 
\nonumber \\
f^{1,-1} = - \frac{e \tau_1 \bm E_{\parallel} \cdot \bm o_+}{2 [1 - i(\omega - \omega_c)\tau_1]} \frac{d f_0(\varepsilon_p)}{d p} \:,
\end{align}
$f^{-1,1} = (f^{1,-1})^*$, and $f^{-1,-1} = (f^{1,1})^*$. The dc electric current is determined by the time-independent asymmetric part of the distribution function described by the harmonics $f^{0,\pm 1}$ for which one obtains
\begin{equation}
f^{0,1} = - \frac{e \tau_1 \zeta_1}{1 - i\omega_c \tau_1} \left(E_z f^{-1,1} + E_z^* f^{1,1} \right) 
\end{equation}
and $f^{0,-1} = (f^{0,1})^*$.

The direct current density $\bm j$ can be then readily found from the general expression
\begin{equation}\label{currenteq}
\bm j^{\rm R} = 2 e \sum_{\bm p} \bm v \left[ f^{0,1} \exp(i \varphi_{\bm p}) + f^{0,-1} \exp(-i \varphi_{\bm p}) \right] \:,
\end{equation}
where the factor $2$ takes into account the spin degeneracy.

Straightforward calculations show that the current density can be presented in the form   
\begin{eqnarray}\label{j_ratchet_phen}
\bm{j}^{\rm R} = L_1 (\bm{E}_{\|} E_z^* + \bm{E}^*_{\|} E_z) 
+ L_2 \, \bm o_z \times (\bm{E}_{\|} E_z^* + \bm{E}_{\|} E_z^* ) \nonumber \\
+ C_1 i (\bm{E}_{\|} E_z^*- \bm{E}^*_{\|} E_z) 
+ C_2 \, \bm o_z \times i (\bm{E}_{\parallel} E_z^* + \bm{E}^*_{\parallel} E_z) \,,\;
\end{eqnarray}
where $\bm o_z$ is the unit vector along the $z$ axis. The coefficients $L_1$, $L_2$, $C_1$, and $C_2$ for a degenerate electron gas are given by
\begin{align}\label{LLCC}
L_1 & =- \dfrac{e^3 N_e}{2 m_c} \dfrac{\zeta_1 \tau_1^2}{1+\omega_c ^2 \tau_1^2} \sum_{s=\pm 1} \dfrac{1-s\omega_c (\omega+s\omega_c)\tau_1^2}{1+(\omega+s\omega_c)^2 \tau_1^2}  \:, \nonumber \\
L_2 & =- \dfrac{e^3 N_e}{2 m_c}  \dfrac{\zeta_1 \tau_1^2}{1+\omega_c ^2 \tau_1^2} \sum_{s=\pm 1} 
\dfrac{(s\omega - 2 \omega_c)\tau_1}{1+(\omega-s\omega_c)^2 \tau_1^2} \:,\nonumber\\
C_1 & = -\dfrac{e^3 N_e}{2 m_c} \dfrac{\zeta_1 \tau_1^2}{1+\omega_c ^2 \tau_1^2} 
\sum_{s=\pm 1} \dfrac{(\omega+2 s\omega_c)\tau_1}{1+(\omega+s\omega_c)^2 \tau_1^2} \:,\nonumber\\
C_2 & = -\dfrac{e^3 N_e}{2 m_c} \dfrac{\zeta_1 \tau_1^2}{1+\omega_c ^2 \tau_1^2} 
\sum_{s=\pm 1} \dfrac{s-\omega_c(\omega+s\omega_c)\tau_1^2}{1+(\omega+s\omega_c)^2 \tau_1^2} \:,
\end{align}
where $N_e = p_F^2/(2 \pi \hbar^2)$ is the electron density, $p_F$ is the Fermi momentum, and $\omega_c$, $\tau_1$, and $\zeta_1$ are taken at the Fermi level. 

The coefficients $L_1$ and $L_2$ describe the electric current excited by a linearly polarized ac electric field (linear ratchet current) whereas $C_1$ and $C_2$ characterize the current contribution which is induced by an elliptically or circularly polarized field and has opposite directions for the right-handed and left-handed polarized radiation (circular ratchet current). In accordance with general symmetry arguments, $C_1$ and $L_1$ describing the current component along $\bm E_{\parallel}$ are even functions of the static magnetic field $B_z$ while $C_2$ and $L_2$ describing the perpendicular component of the current are odd functions of $B_z$.

\begin{figure} 
\includegraphics[width=0.9\columnwidth]{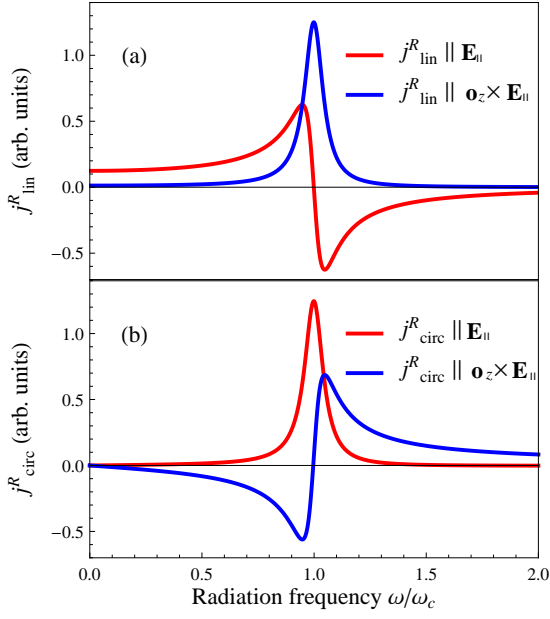}
\caption{\label{pge} 
Dependences of the projections of the linear ratchet current (panel a) and projections of the circular ratchet current (panel b) on the electric field frequency $\omega$. Curves are calculated after Eqs.~\eqref{j_ratchet_phen} and~\eqref{LLCC} for $\omega_c \tau_1=20$. 
} 

\end{figure}

The dependences of the linear ratchet current and the circular ratchet current 
on the field frequency $\omega$ are shown in Figs.~\ref{pge}a and~\ref{pge}b, respectively. At zero frequency, only the linear ratchet current is generated. Its direction in the QW plane with respect to $\bm E_{\parallel}$ is determined by the ratio $(L_2/L_1)_{\omega=0} = -2\omega_c \tau_1 /[1 - (\omega_c \tau_1)^2]$.
The deflection of the current direction from $ \bm E_{\parallel}$ is caused by the Lorentz force acting upon the electrons.
The circular ratchet effect emerges at finite frequencies of the ac electric field. The magnitude of the circular ratchet current is proportional to $\omega$ at small $\omega$, the current direction is determined by the ratio $C_2/C_1 = - \omega_c \tau_1 [3-(\omega_c\tau_1)^2] /[1 - 3(\omega_c \tau_1)^2]$. At the cyclotron resonance conditions, both the linear and the circular ratchet currents drastically increase. Their directions in the QW plane are very sensitive to the frequency detuning $\omega - \omega_c$, see Figs.~\ref{pge}a and~\ref{pge}b. Finally, far from the cyclotron resonance, the high-frequency asymptotic behavior of the ratchet currents is described by
$L_1 \propto 1/\omega^2$, $L_2 \propto 1/\omega^4$, and $C_1, C_2 \propto 1/\omega$.

The magnitude of the ratchet current can be estimated from Eqs.~\eqref{j_ratchet_phen} and~\eqref{LLCC}.  For the 
electric field amplitude $E=10$~kV/cm, the static magnetic field $B=4$~T, the momentum relaxation time $\tau_1=10^{-12}$~s,
the effective mass $m_c=0.07 m_0$ (corresponding to GaAs-based QWs), the carrier density $N_e=2 \cdot 10^{11}$~cm$^{-2}$, the QW width
$d=10$~nm, and the degree of QW structure inversion asymmetry $ \langle \operatorname{Re} V^*_{11}(\bm p, \bm p') V_{12} (\bm p, \bm p') \rangle/ \langle | V_{11} (\bm p, \bm p') |^2\rangle=0.1$, an estimation gives $j^{\rm R} \sim 2$~$\mu$A/cm for the ratchet current at the cyclotron resonance.

\section{Tilted magnetic field}

Now we consider the geometry of a tilted static magnetic field $\bm B$ and ac electric field $\bm E(t)$ polarized in the QW plane. 
In this case, an asymmetry of the electron distribution in the momentum space and, hence, a dc electric current may emerge due to
the asymmetry of electron scattering induced by the in-plane component $\bm B_{\parallel}$ of the magnetic field. Such a magnetic quantum ratchet effect in a purely in-plane magnetic field was theoretically studied in Refs.~\onlinecite{Falko1989,Tarasenko2011} and has been recently observed in graphene.~\cite{Drexler2013} Here, we develop a microscopic theory of the effect for a tilted magnetic field where the in-plane component $\bm B_{\parallel}$ induces the scattering asymmetry while the out-of-plane component $B_z$ causes the cyclotron motion of the electrons and leads to a resonant enhancement of the ratchet current.   

To first order in $\bm B_{\parallel}$, the rate of elastic electron scattering can be generally presented in the form
\begin{equation}\label{W2_pp}
W_{\bm{p} \bm{p}'} = W_{\bm{p} \bm{p}'}^{(0)} + e B_{\parallel} \: W_{\bm{p} \bm{p}'}^{(1)} \:,
\end{equation}
where $W_{\bm{p} \bm{p}'}^{(0)}$ is the scattering rate at zero field and $W_{\bm{p} \bm{p}'}^{(1)}$ is the field induced correction; $W_{\bm{p}, \bm{p}'}^{(0)} = W_{- \bm{p}', -\bm{p}}^{(0)}$ and $W_{\bm{p}, \bm{p}'}^{(1)} = - W_{- \bm{p}', -\bm{p}}^{(1)}$ due to time inversion symmetry. Microscopically, the scattering asymmetry stems from the quantum analogue of the Lorentz force which pushes moving electrons to the top or bottom interfaces of the QW. For a QW structure with the simple parabolic energy spectrum, the term $W_{\bm{p} \bm{p}'}^{(1)}$ is given by~\cite{Tarasenko08} 
\begin{multline}\label{magneticw1}
W_{\bm{p} \bm{p}'}^{(1)} = - \dfrac{4 \pi}{\hbar m_e c} 
\left[ \frac{B_x}{B_{\parallel}}(p_y + p'_y) - \frac{B_y}{B_{\parallel}}(p_x + p'_x)  \right] \\
 \sum\limits_{\nu \neq 1 } \dfrac{z_{\nu 1}}{\varepsilon_{ \nu 1}} \langle \operatorname{Re} V^*_{11}(\bm p, \bm p') V_{1\nu} (\bm p, \bm p')  \rangle  \, \delta (\varepsilon_p -\varepsilon_{p'}) \,,
\end{multline}
where $m_e$ is the effective mass.

To calculate the dc current due to the magneto-induced ratchet effect we solve the Boltzmann equation~\eqref{kinetic_equation} with the scattering rate~\eqref{W2_pp}. In the harmonics representation, the Boltzmann equation assumes the form of the linear equation set
\begin{multline}
\label{linear_set2}
\Gamma^{n,m} f^{n,m} + e B_{\|}\sum\limits_{l} \left(u_{0,l} f^{n,m-l} -u_{m,l} f^{n,-l}\right)\\+
\dfrac{e \bm{E}_{\|}}{2 } \cdot
\left(\bm{o}_{-} \hat{K}^m_{-} f^{n-1,m-1} +
\bm{o}_{+} \hat{K}^m_{+} f^{n-1,m+1}\right)\\
+\dfrac{e \bm{E}_{\|}^*}{2} \cdot 
\left(\bm{o}_{-} \hat{K}^m_{-} f^{n+1,m-1} +
\bm{o}_{+} \hat{K}^m_{+} f^{n+1,m+1}\right)=0
\:,
\end{multline}
where $u_{n,m}$ are defined by
\begin{equation}
u_{n,m} =  \int \dfrac{d \varphi_p}{2 \pi} \sum\limits_{\bm{p}'}  W_{\bm{p} \bm{p}'}^{(1)} e^{-i n \varphi_{\bm{p}}-i m\varphi_{\bm{p}'}}\:.
\end{equation}
The harmonics $u_{n,m}$ satisfy the conditions $u_{n,m}=u_{-n,-m}^*$ and $u_{n,m}=(-1)^{n+m+1} u_{m,n}$ due to the reality of the scattering rate and time inversion symmetry, respectively. It also follows from Eq.~\eqref{magneticw1} that, for central scattering, the only nonzero harmonics are $u_{n,n\pm1}$.

We seek a solution of the equation set~\eqref{linear_set2} in the form of a perturbation series in $\bm E$ and $\bm B_{\parallel}$.
To first order in the electric field amplitude, one obtains the harmonics $f^{\pm1,\pm 1}$ which are given by Eqs.~\eqref{f11}. The second iteration in $\bm{E}_{\|}$ and $\bm B_{\|}$ yields
\begin{align}
f^{1,2} &= \dfrac{\tau_2 eB_{\|} (u_{2,-1}-u_{0,1})f^{1,1}}{1-i ( \omega_c +2\omega_c)\tau_2}\:, \\
f^{1,-2} &= \dfrac{\tau_2 eB_{\|} (u_{-2,1}-u_{0,-1})f^{1,-1}}{1-i ( \omega_c -2\omega_c)\tau_2}\:,\nonumber\\
f^{0,2} &= -\dfrac{e \tau_2[ (\bm{E}_{\|} \cdot \bm{o}_-)\hat{K}^2_{-}f^{-1,1}+(\bm{E}_{\|}^* \cdot \bm{o}_-)\hat{K}^2_{-}f^{1,1}]}{2 (1-2 i \omega_c \tau_2)} \nonumber \:,
\end{align}
and $f^{-n,-m}=(f^{n,m})^*$. The dc electric current is determined by the harmonics $f^{0,\pm1} \propto E^2 B_{\parallel}$; they have the form
\begin{align}
f^{0,1} = & -\dfrac{e \tau_1[ (\bm{E}_{\|} \bm{o}_+)\hat{K}^1_{+}f^{-1,2}+(\bm{E}_{\|}^* \bm{o}_+)\hat{K}^1_{+}f^{1,2}]}{2 (1-i \omega_c \tau_1)}  \\
&+ \dfrac{\tau_1 e B_{\|}(u_{1,-2}-u_{0,-1})f^{0,2}}{1-i \omega_c \tau_1} \: \nonumber
\end{align}
and $f^{0,-1}=(f^{0,1})^*$.

Finally, calculating the dc electric current $\bm j^{\rm R}$ following the general expression~\eqref{currenteq} we obtain
\begin{align}\label{j_M}
j_x^{\rm R} & = \left(\operatorname{Re} M_0 + 
\xi_1\operatorname{Re} M_L + \xi_2 \operatorname{Im} M_L + \xi_3 \operatorname{Re} M_C \right) E^2 B_{\parallel} , \nonumber \\
j_y^{\rm R} & =- \left(\operatorname{Im} M_0 +
\xi_1 \operatorname{Im} M_L - \xi_2 \operatorname{Re} M_L + \xi_3 \operatorname{Im} M_C \right) E^2 B_{\parallel} , 
\end{align}
where $\xi_1=(|E_x|^2-|E_y|^2)/|E|^2$, $\xi_2=(E_x E_y^*+E_x^* E_y)/|E|^2$, and $\xi_3=i(E_x E_y^*-E_x^* E_y)/|E|^2$ are the Stokes parameters determining the polarization state of the ac electric field. Accordingly, the parameters $M_0$, $M_L$, and $M_C$ describe the magnitudes of the magneto-induced ratchet effect independent of the ac field polarization, the linear magneto-induced ratchet (LMR) effect, and the circular magneto-induced ratchet (CMR) effect, respectively. For a degenerate electron gas, the parameters are given by 
\begin{align}\label{M}
M_0 &= \frac{e^4 N_e \gamma^* p_F}{4 m_e^2} \left(\dfrac{1}{\Gamma^{0,1} }\right)'_{\varepsilon_p}  
\left(\dfrac{1}{\Gamma^{1,2}\Gamma^{1,1}} + \dfrac{1}{\Gamma^{-1,2}\Gamma^{-1,1}} \right) , \nonumber \\
M_L &= -\dfrac{e^4 N_e}{4 m_e^2 \,p_F^2} \left(\dfrac{\gamma \, p^3}{\Gamma^{0,1}\Gamma^{0,2}} \right)'_{\varepsilon_p}
\left(\dfrac{1}{\Gamma^{1,1}} + \dfrac{1}{\Gamma^{-1,1}} \right) , \nonumber \\
M_C &= \dfrac{e^4 N_e \gamma^{*} p_F}{ 4 m_e^2} \left(\dfrac{1}{\Gamma^{0,1}}\right)'_{\varepsilon_p}  \left(\dfrac{1}{\Gamma^{1,2}\Gamma^{1,1}} - \dfrac{1}{\Gamma^{-1,2}\Gamma^{-1,1}} \right) \:, 
\end{align}
where $\gamma=u_{-2,1}-u_{0,-1}$ is the coefficient describing the electron scattering anisotropy induced by the in-plane component 
of the magnetic field. All the values in Eq.~\eqref{M} are taken at the Fermi level. We note that an additional contribution to the
polarization independent current may arise due to asymmetry of the energy relaxation of hot carriers in the magnetic field.\cite{Tarasenko08,Kibis1999} This current depends on the details of election-phonon interaction and is out of the scope of the present paper.
  
\begin{figure} 
\includegraphics[width=1\columnwidth]{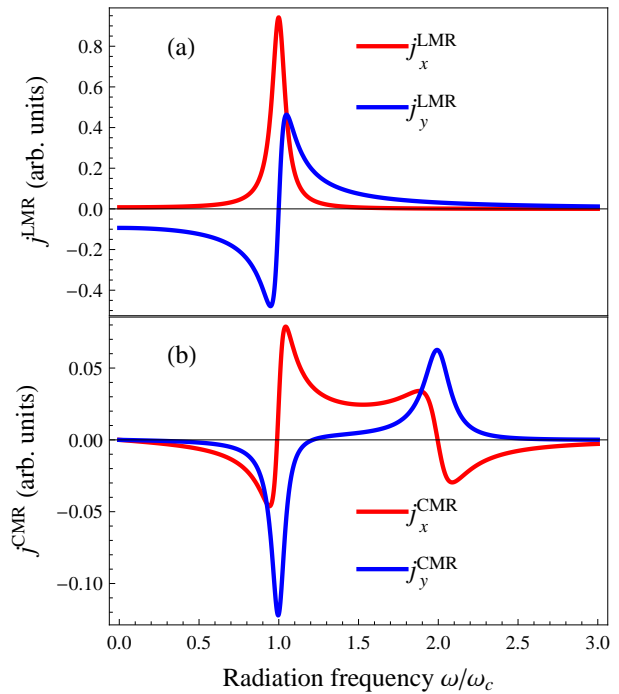}
\caption{\label{mpge} 
Dependences of the projections of the linear magneto-induced ratchet current (panel a) and projections of the circular magneto-induced ratchet current (panel b) on the electric field frequency $\omega$. Curves are calculated after Eqs.~\eqref{j_M} and~\eqref{M} for $\omega_c \tau_1(E_F)=20$, $\tau_1(\varepsilon_p) = 2 \tau_2(\varepsilon_p) \propto \varepsilon_p$, and real and energy independent $\gamma$.
} 
\end{figure}

Figure~\ref{mpge} shows the frequency dependence of the LMR and CMR currents which are determined by $M_L(\omega)$ and $M_R(\omega)$, respectively. The curves are calculated for the case when the electron mobility is limited by Coulomb impurities, 
$\tau_1(\varepsilon_p) = 2 \tau_2(\varepsilon_p) \propto \varepsilon_p$, and the intersubband scattering is determined by short-range defects, i.e., $\gamma$ is independent of the electron energy. The coefficient $\gamma$ is assumed to be real which corresponds to the
geometry $\bm B_{\parallel} \parallel y$. One can see that both the LMR and CMR currents drastically increase at cyclotron resonance. However, 
the resonant contribution to the LMR current exceeds the resonant contribution to CMR current by the factor of $\omega_c \tau_1$.
Moreover, the CMR effect emerges due to the energy dependence of the momentum relaxation time. The CMR current has an additional resonance at $\omega = 2 \omega_c$ which occurs in the spectral range where the Drude absorbance has no peculiarity. In the quantum-mechanical picture, this additional resonance corresponds to the transitions between the Landau levels $n$ and $n+2$.

\section{Current distribution in the quantum well plane}

In experiments, ratchet currents can be excited by the ac electric field of terahertz or microwave radiation focused on the sample~\cite{Drexler2013,Belkov2008,Olbrich2013,Dantscher2015}.
The dc electric signal can be measured via the voltage drop across contacts in the open-circuit configuration where the  
net electric current in the circuit is vanishingly small. The voltage drop is defined by the distribution of the electrostatic potential $\Phi(x,y)$ in the QW plane. The latter is determined by the spatial distribution of the radiation-induced ratchet current $\bm j^{\rm R}(x,y)$, the drift current $\bm j^{\rm DR}(x,y)$, which tends to compensate the ratchet current, and the boundary conditions at the sample edges. 

The continuity equation requires
\begin{equation}\label{cont}
\nabla \cdot (\bm j^{\rm R} + \bm j^{\rm DR} ) = 0 \:.
\end{equation}
The components of the drift current are given by 
\begin{equation}
j_{\alpha}^{\rm DR} = - \sum_{\beta} \sigma_{\alpha\beta} \nabla_{\beta} \Phi \:,
\end{equation}
where $\sigma_{\alpha\beta}$ is the conductivity tensor in the magnetic field, 
\begin{align}\label{cond}
\sigma_{xx} = \sigma_{yy} = \frac{\sigma_0}{1+(\omega_c\tau_1)^2} , \nonumber \\
\sigma_{xy} =-\sigma_{yx} = \frac{\omega_c \tau_1 \, \sigma_0 }{1+(\omega_c\tau_1)^2} ,
\end{align}
$\sigma_0 = N_e e^2 \tau_1 / m_e$ is the conductivity at zero magnetic field.

From Eqs.~\eqref{cont}-\eqref{cond} we obtain the Poisson equation
\begin{equation}\label{poison}
\Delta \Phi = (\nabla \cdot \bm j^{\rm R}) / \sigma \:,
\end{equation}
where $\sigma=\sigma_0/(1+\omega_c^2 \tau^2)$. Equation~\eqref{poison} should be solved with the boundary conditions. For finite-size samples, we consider the boundary condition of zero total electric current flowing across the sample edges, $\bm j_{\bm n}^{\rm R} + \bm j_{\bm n}^{\rm DR} = 0$. 

Solution of the Eq.~\eqref{poison} can be generally presented in the form
\begin{equation}
\Phi(\bm{r})=\int \bm{j}^{\rm R}(\bm{r}') \cdot \bm{\rho} (\bm{r},\bm{r}') d \bm{r}'\:,
\end{equation}
where $\bm{\rho} (\bm{r},\bm{r}')$ can be interpreted as the function of nonlocal resistance. The explicit form of the function 
$\bm{\rho} (\bm{r},\bm{r}')$ depends on the sample shape and, in general case, can be calculated numerically. Below we discuss the spatial distribution of the electrostatic potential in infinite, semi-infinite, and rectangular-shape structures and present some analytical results.
 
For an infinite two-dimensional system, the solution of the two-dimensional Poisson equation can be readily found by 
Green's function method, which yields
\begin{equation}
\label{greenfunction}
\bm{\rho} (\bm{r},\bm{r}') = -\dfrac{1}{2\pi \sigma} \dfrac{\bm{r}-\bm{r}'}{(\bm{r}-\bm{r}')^2}\:.
\end{equation}
The magnetic field $B_z$ only scales the function $\bm{\rho} (\bm{r},\bm{r}')$. However the field also affect the magnitude and direction of $\bm{j}^{\rm R}(\bm{r}')$ and, hence, the spatial distribution of the electrostatic potential.

Using the method of mirror images and the function $\bm{\rho} (\bm{r},\bm{r}')$ for the infinite system, one can derive an analytical equation for $\bm{\rho} (\bm{r},\bm{r}')$ in a semi-infinite system. For the system $x \geq 0$ with a single boundary at $x=0$, we obtain
\begin{align}\label{greenfunction1}
\rho_x (\bm{r},\bm{r}') &= -\dfrac{1}{2\pi \sigma} \Biggl[\dfrac{x-x'}{(x-x')^2+(y-y')^2} \nonumber \\ 
&- \dfrac{(1-\omega_c^2\tau_1^2)(x+x') - 2\omega_c\tau_1(y-y')}{(1+\omega_c^2\tau_1^2)[(x+x')^2+(y-y')^2]} \Biggr] , \nonumber \\
\rho_y (\bm{r},\bm{r}') &= -\dfrac{1}{2\pi \sigma} \Biggl[\dfrac{y-y'}{(x-x')^2+(y-y')^2} \nonumber \\ 
&+ \dfrac{(1-\omega_c^2\tau_1^2)(y-y') - 2\omega_c\tau_1(x+x')}{(1+\omega_c^2\tau_1^2)[(x+x')^2+(y-y')^2]} \Biggr] . 
\end{align}

By adding three additional boundaries and considering a net consisting of rectangular cells, one can also obtain the function
of nonlocal resistance for a finite-size rectangular structure with the edges at $x = 0, \, a$ and $y = 0, \,b$. The function is given by
\begin{multline}
\rho_x = \dfrac{1}{4\sigma}\sum\limits_{n = -\infty}^{\infty} \Biggl[ \dfrac{(1/b)\sinh(\pi r'_-/b)}{\cosh(2 \pi n a/b- \pi r_-/b)
-\cosh(\pi r'_-/b)} \\
+ \dfrac{1-\omega_c^2 \tau_1^2}{1 + \omega_c^2 \tau_1^2} \dfrac{ (1/b) \sinh(\pi r'_+/b )}{\cosh(2 \pi n a/b - \pi r_-/b)
-\cosh(\pi r'_+/b)} \\
+ \dfrac{2 \omega_c \tau_1}{1 + \omega_c^2 \tau_1^2}
\dfrac{(1/a) \sinh(i \pi r'_-/a)}{\cosh(2 \pi n b/a + i \pi r_+ /a ) - \cosh(i \pi r'_-/a)} \Biggr] + \text{c.c.} \:,
\end{multline}
where $r_{\pm} = x \pm i y$, $r'_{\pm} = x' \pm i y'$; the function $\rho_y$ is obtained from $\rho_x$ by replacing $x$, $y$, 
$x'$ $y'$, $a$, and $b$ with $y$, $-x$, $y'$, $-x'$, $b$, and $a$, respectively.

\begin{figure}[t]
\includegraphics[width=0.9\columnwidth]{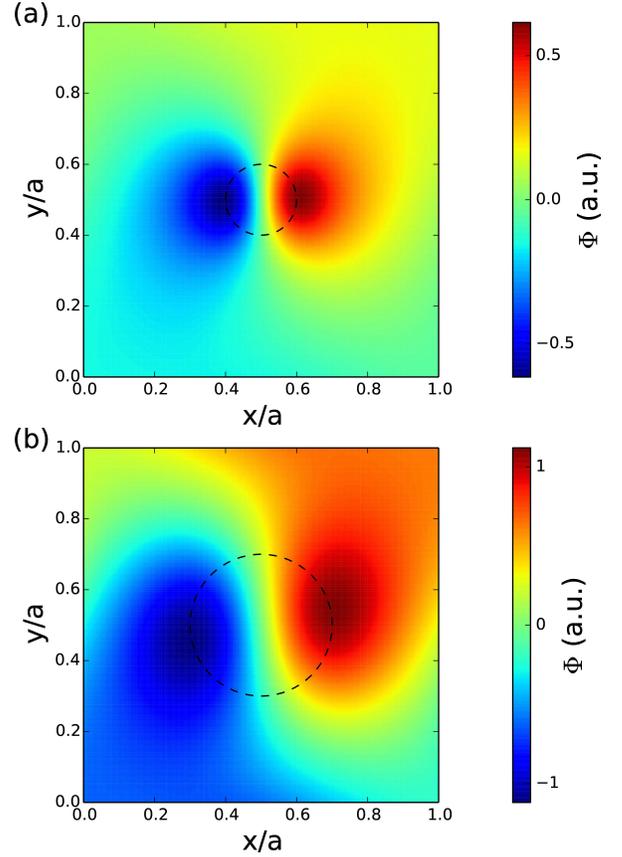}
\caption{\label{potential_distribution} 
Spatial distributions of the electrostatic potential induced by a ratchet current generated in the center area of a square-shape structure in a magnetic field. Distributions are calculated for ${\omega_c\tau_1 = 2}$ and the Gaussian distribution of the ratchet current density given by Eq.~\eqref{gaussian} for $D/a = 0.1$ (figure a) and $D/a = 0.2$ (figure b). Dashes circles sketch the areas
of the ratchet current generation.}
\end{figure}

Figure~\ref{potential_distribution} demonstrates the spatial distribution of the electrostatic potential $\Phi(x,y)$ in a square-shape structure when the ratchet current is generated in the central area of the structure. The spatial distribution of the
ratchet current density is taken in the form of the Gaussian function
\begin{equation}\label{gaussian}
\bm{j}^{\rm R} = \bm{j}_0^{\rm R} \exp\left(-\dfrac{(x-a/2)^2+(y-a/2)^2}{D^2}\right) \:,
\end{equation}
where $\bm{j}_0^{\rm R} \parallel x$ and $a$ is the structure size. The potential $\Phi$ reaches extremal values at the positions close to the border of the current generation spot while at the sample edges it decreases. The magnetic field changes the potential amplitude
and also twists the potential spatial distribution that can be seen from equipotential lines. The calculation shows that the voltage drop
between the points of the highest and lowest potential is about $10$~mV for the spot radius $D=1$~mm, the ratchet current magnitude
$\bm{j}_0^{\rm R} = 2$~$\mu$A/cm and the parameters of the structure presented in Sec.~\ref{Sec_perpB}. The calculations also reveal that experimentally measured voltage drops across contacts may drastically depend on the contact positions in the structure and the structure geometry.

\section{Summary}

We have developed a microscopic theory of the ratchet transport of electrons confined in a quantum well. It has been shown that
the magnitude of the direct electric current induced by an alternating electric field is increased in an external static magnetic field at the cyclotron resonance when the electric field frequency matches the cyclotron frequency. The magnetic field gives rise also to an additional mechanism of the ratchet current generation which stems from an asymmetry of electron scattering in the magnetic field. The magneto-induced ratchet effect has a resonant behavior both at the cyclotron resonance and its first subharmonic, with the 
current being sensitive to the electric field polarization and the mechanism of electron scattering. We have also analyzed the spatial distribution of the electrostatic potential induced by the ratchet current and shown that the voltage drop between 
contacts is highly sensitive to the sample geometry and the contact positions. The resonant behavior of the ratchet current can be used for the study of electron scattering mechanisms and the development of tunable fast detectors of microwave and terahertz radiation.

\begin{acknowledgments}
This work has been supported by the Russian Science Foundation (project no. 14-12-01067). GVB also acknowledges support from the ''Dynasty'' Foundation.
\end{acknowledgments}


\begin{thebibliography}{99}

% FETs

\bibitem{Dyakonov1996} M. I. Dyakonov and M. S. Shur, 
Detection, mixing, and frequency multiplication of terahertz radiation by two-dimensional electronic fluid,
IEEE Trans. Electron Devices {\bf 43}, 380 (1996).

\bibitem{Knap2009} W. Knap, M. Dyakonov, D. Coquillat, F. Teppe, N. Dyakonova, J. Lusakowski, K. Karpierz, M. Sakowicz, G. Valusis, 
D. Seliuta, I. Kasalynas, A. El Fatimy, Y. M. Meziani, and T. Otsuji, J. Infrared Millim,
Field effect transistors for terahertz detection: Physics and first imaging applications,
Terahertz Waves {\bf 30}, 1319 (2009).

% Asymmetric lattices

\bibitem{Blanter1998} Ya. M. Blanter and M. B\"{u}ttiker,
Rectification of fluctuations in an underdamped ratchet,
\prl {\bf 81}, 4040 (1998).

\bibitem{Hoehberger2001} E. M. H\"{o}hberger, A. Lorke, W. Wegscheider, and M. Bichler,
Adiabatic pumping of two-dimensional electrons in a ratchet-type lateral superlattice,
Appl. Phys. Lett. {\bf 78}, 2905 (2001).

\bibitem{Olbrich2009} P. Olbrich, E. L. Ivchenko, R. Ravash, T. Feil, S. D. Danilov, J. Allerdings, D. Weiss,
D. Schuh, W. Wegscheider, and S. D. Ganichev,
Ratchet effects induced by terahertz radiation in heterostructures with a lateral periodic potential,
\prl {\bf 103}, 090603 (2009).

\bibitem{Popov2013} V. V. Popov,
Terahertz rectification by periodic two-dimensional electron plasma,
Appl. Phys. Lett. {\bf 102}, 253504 (2013).

\bibitem{Budkin2014} G. V. Budkin and L. E. Golub,
Orbital magnetic ratchet effect,
\prb {\bf 90}, 125316 (2014).

\bibitem{Rozhansky2015} I. V. Rozhansky, V. Yu. Kachorovskii, M. S. Shur,
Helicity-driven ratchet effect enhanced by plasmons,
\prl {\bf 114}, 246601 (2015).

% Asymmetric antidots

\bibitem{Entin2006} M. V. Entin and L. I. Magarill,
Photocurrent in nanostructures with asymmetric antidots: Exactly solvable model,
\prb {\bf 73}, 205206 (2006).

\bibitem{Sassine08} S. Sassine, Yu. Krupko, J.-C. Portal, Z. D. Kvon, R. Murali, K. P. Martin,
G. Hill, and A. D. Wieck, 
Experimental investigation of the ratchet effect in a two-dimensional electron system with broken spatial inversion symmetry,
\prb {\bf 78}, 045431 (2008).

\bibitem{Chepelianskii08} A. D. Chepelianskii, M. V. Entin, L. I. Magarill, and D. L. Shepelyansky,
Ratchet transport of interacting particles,
\pre {\bf 78}, 041127 (2008).

\bibitem{Bisotto2011} I. Bisotto, E. S. Kannan, S. Sassine, R. Murali, T. J. Beck, L. Jalabert, and J.-C. Portal,
Microwave based nanogenerator using the ratchet effect in Si/SiGe heterostructures,
Nanotechnology {\bf 22}, 245401 (2011).

\bibitem{Koniakhin2014} S. V. Koniakhin,
Ratchet effect in graphene with trigonal clusters,
EPJ B {\bf 87}, 216 (2014).

% Macroscopically homogenious structures

\bibitem{Sturman_book} B.I.~Sturman and V.M.~Fridkin, 
\textit{The Photovoltaic and Photorefractive Effects in
Non-Centrosymmetric Materials} (Gordon and Breach Science
Publishers, New York, 1992).


\bibitem{Falko1989} V. I. Fal'ko, 
Rectifying properties of 2D inversion layers in a parallel magnetic field,
Fiz. Tvedr. Tela {\bf 31}, 29 (1989) [Sov. Phys. Solid State {\bf 31}, 561 (1989)].

\bibitem{Tarasenko2011} S.A.~Tarasenko,
Direct current driven by ac electric field in quantum wells, 
\prb {\bf 83}, 035313 (2011).

\bibitem{Entin2013} M. V. Entin and L. I. Magarill,
Photogalvanic current in a parabolic well,
Pis'ma Zh. Eksp. Teor. Fiz. {\bf 97}, 737 (2013) [JETP Lett. {\bf 97}, 639 (2013)].
 

\bibitem{Drexler2013} C. Drexler, S.A. Tarasenko, P. Olbrich, J. Karch, M. Hirmer, F. M\"{u}ller, M. Gmitra, J. Fabian, R. Yakimova, S. Lara-Avila, S. Kubatkin, M. Wang, R. Vajtai, P.M. Ajayan, J. Kono, and S.D. Ganichev, 
Magnetic quantum ratchet effect in graphene, 
Nature Nanotechnol. {\bf 8}, 104 (2013).

% Symmetry and spectrum studies

\bibitem{Belkov2008} V. V. Bel'kov, P. Olbrich, S. A. Tarasenko, D. Schuh, W. Wegscheider, T. Korn, Ch. Sch\"{u}ller, D. Weiss, W. Prettl, and S.D. Ganichev, 
Symmetry and spin dephasing in (110)-grown quantum wells, 
Phys. Rev. Lett. {\bf 100}, 176806 (2008).

\bibitem{Olbrich2013}	P. Olbrich, C. Zoth, P. Vierling, K.-M. Dantscher, G. V. Budkin, S. A. Tarasenko, V. V. Bel'kov, D. A. Kozlov, Z. D. Kvon, N. N. Mikhailov, S. A. Dvoretsky, and S. D. Ganichev, 
Giant photocurrents in a Dirac fermion system at cyclotron resonance, 
\prb {\bf 87}, 235439 (2013).

\bibitem{Dantscher2015} K.-M. Dantscher, D. A. Kozlov, P. Olbrich, C. Zoth, P. Faltermeier, M. Lindner, G. V. Budkin, S. A. Tarasenko, V. V. Belkov, Z. D. Kvon, N. N. Mikhailov, S. A. Dvoretsky, D. Weiss, B. Jenichen, S. D. Ganichev,
Cyclotron resonance assisted photocurrents in surface states of a 3D topological insulator based on a strained high mobility HgTe film,
arXiv:1503.06951. 

% Detectors

\bibitem{Ganichev2007} S. D. Ganichev, J. Kiermaier, W. Weber, S. N. Danilov, D. Schuh, Ch. Gerl, W. Wegscheider, W. Prettl, D.
Bougeard, and G. Abstreiter,
Subnanosecond ellipticity detector for laser radiation,
Appl. Phys. Lett. {\bf 91}, 091101 (2007).

% CR

\bibitem{Dmitriev1991} A. P. Dmitriev, S. A. Emelyanov, S. V. Ivanov, P. S. Kop'ev, Ya. V. Terent'ev, and I. D. Yaroshetsky, 
Drag photocurrent in a 2D electron gas near the cyclotron resonance and its first subharmonic,
Pis'ma Zh. Eksp. Teor. Fiz. {\bf 54}, 460 (1991) [JETP Lett. {\bf 54}, 462 (1991)].

\bibitem{Stachel2014} S. Stachel, G. V. Budkin, U. Hagner, V. V. Bel'kov, M. M. Glazov, S. A. Tarasenko, S. K. Clowes, T. Ashley, A. M. Gilbertson, S. D. Ganichev, 
Cyclotron-resonance-assisted photon drag effect in InSb/InAlSb quantum wells excited by terahertz radiation, 
\prb {\bf 89}, 115435 (2014).

% Scattering asymmetry in B-field

\bibitem{Tarasenko08}  S.A.~Tarasenko,
Electron scattering in quantum wells subjected to an in-plane magnetic field,
\prb {\bf 77}, 085328 (2008).

\bibitem{Kibis1999} O.V.~Kibis, Novel effects of electron-phonon interaction in quasi-two-dimensional structures
located in a magnetic field, Sov. Phys. JETP {\bf 88}, 527 (1999) 
[Zh. Eksp. Teor. Fiz. {\bf 115}, 959 (1999)].


\end{thebibliography}
\end{document}